# Giant Domain-Wall Hall Magnetoresistance in Magnetic Topological Semimetal

Jinying Yang[1,2], Qingqi Zeng[1,3], Yibo Wang[1,2], Meng Lyu[1], Yang Liu[1,2], Xingchen Liu[1,2], Xuebin Dong[1,2], Binbin Wang[1], Xiyang Li[1], Enke Liu[1,2,*]

[1]Beijing National Laboratory for Condensed Matter Physics, Institute of Physics, Chinese Academy of Sciences, Beijing 100190, China
[2]School of Physical Sciences, University of Chinese Academy of Sciences, Beijing 100049, China
[3]Guangdong Provincial Key Laboratory of Optical Information Materials and Technology, Institute for Advanced Materials, South China Academy of Advanced Optoelectronics, South China Normal University, Guangzhou, 510006, China

[*]Email: ekliu@iphy.ac.cn

## Abstract

Magnetic topological semimetals exhibit emerging magneto-transport behaviors, such as the giant anomalous Hall effect (AHE), chiral Hall effect, and antisymmetric magnetoresistance. In this work, based on the magnetic Weyl semimetal $Co_3Sn_2S_2$, we report an intriguing longitudinal domain-wall Hall magnetoresistance in multi-domain states. According to a multi-domain model, a concise formula of this Hall magnetoresistance was revealed and verified experimentally. Rather than the real change of longitudinal resistance, this Hall magnetoresistance originates from an additional electric field distribution induced by the transverse giant AHE through the domain wall, which can be directly correlated to the Berry phase of topological Weyl bands. In $Co_3Sn_2S_2$ devices, the Hall magnetoresistance was an order of magnitude larger than that of conventional magnetic materials, indicating its potential for multi-resistance-state modulation via the Weyl-enhanced AHE.

The combination of magnetism and topology has brought and enhanced several exotic phenomena [1-4]. The topological transport of magnetic Weyl semimetals in low dimensions is important for device physics and spintronics applications, as well as for the potential quantum AHE in the two-dimensional limit [5]. As a well-established magnetic Weyl semimetal [1,6-8], $Co_3Sn_2S_2$ is an ideal platform for various interesting effects and phenomena. The large Berry curvature from the Weyl nodes and nodal lines around the Fermi level results in giant anomalous Hall/Nernst effects and exotic transports, such as the antisymmetric magnetoresistance [9] and thermoelectric effects [10].

In magnetic Weyl semimetals, topological transport effects modulated by spin structures have attracted considerable attention. Theoretically, an axial electromagnetic field can arise with the magnetic texture, which induces charge pumping without Joule heating [11]. The velocity of the domain wall is expected to be an order of magnitude higher than those of conventional magnetic metals [12]. Owing to their excellent performances and expectations, magnetic Weyl semimetals exhibit great potential for spintronics applications. Therefore, investigating the transport properties of magnetic Weyl semimetals with magnetic domains and domain walls is meaningful.

In this work, we study the transport properties of the magnetic Weyl semimetal $Co_3Sn_2S_2$ nanoflakes in the multi-domain states. A large resistance, antisymmetric in magnetization and magnetic field, appears at low magnetic fields near the Curie temperature. This signal dose not overlap during cooling and warming under the same magnetic field. Magnetic field-dependent experiments demonstrate that an anomalous Hall-like characteristic appears in the longitudinal resistance. Further studies found that this signal originates from the contribution of the AHE in the multi-domain states, rather than from the real longitudinal resistance of sample. While referring to this resistance as domain-wall Hall magnetoresistance, a model and calculation formula were proposed to understand and calculate it.

The chemical vapor transport method was employed to grow $Co_3Sn_2S_2$ single crystal nanoflakes. The thicknesses of the samples were characterized by atomic force microscopy. Samples with appropriate thicknesses were selected to fabricate standard Hall bars by electron beam lithography and ion-beam etching. Electron beam evaporation was used to fabricate the electrodes. The schematic of the transport measurements is shown in the inset of Fig. 1(a).

The temperature dependence of longitudinal and transverse Hall resistances was measured under low magnetic fields, as shown in Fig. 1. The Curie temperature ($T_C$) of $Co_3Sn_2S_2$ nanoflake is 177 K, which is consistent with previous reports [1,3,13,14]. As the temperature decreases, a large negative contribution of additional resistance ($\Delta R_{xx}$, ~50 Ω) unexpectedly emerges at a low field of 5 Oe, resulting in a clear dip in the longitudinal resistance between $T_C$ and 155 K (Fig. 1(a)). As the magnetic field increases, the dip gradually disappears, leaving a normal resistance-temperature curve above 20 Oe. Interestingly, the sign of $\Delta R_{xx}$ reverses upon changing the direction of the magnetic field. A positive $\Delta R_{xx}$ plateau observed at low fields also disappears above -20 Oe. The $\Delta R_{xx}$ component is apparently antisymmetric in the magnetic field. This significant resistance anomaly does not originate from the Hall contribution owing to the possible mismatch of the measuring electrodes, because it only appears in a narrow temperature range. As shown in Fig. 1(b), the anomalous Hall response was measured below $T_C$. Even at extremely low fields, the anomalous Hall resistance is very large. In the same temperature range ($T_C$–155 K), the values of $R_{yx}$ are relatively low below 20 Oe where the $\Delta R_{xx}$ plateau exists. The $R_{yx}$ increases and they overlap together along with the vanishing of the $\Delta R_{xx}$ plateau in higher fields. Similar behavior occurs in the case of negative magnetic fields. The Hall results indicate that the sample is in a multi-domain state below 20 Oe in the range between $T_C$ and 155 K and it becomes almost saturated as a single magnetic domain above 50 Oe, which implies that the $\Delta R_{xx}$ component just exists in the multi-domain state. Below 155 K, this unusual behavior disappears even at low fields because the magnetic anisotropy of the sample increases and the single-domain state forms spontaneously.

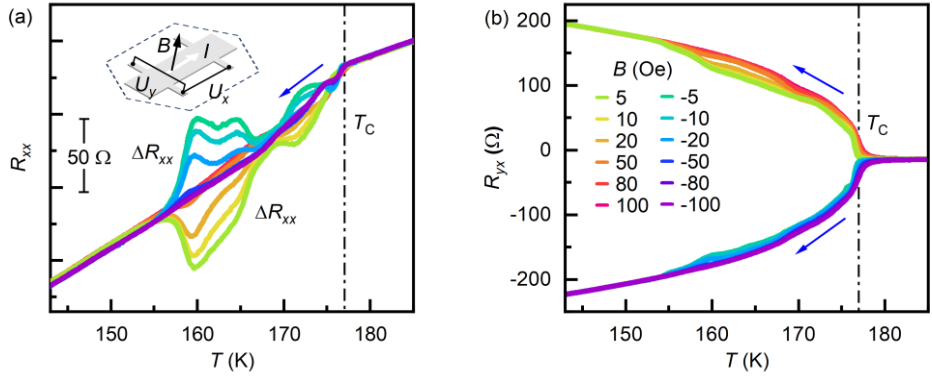

FIG. 1. Temperature-dependent longitudinal and transverse Hall resistances of the $Co_3Sn_2S_2$ nanoflake. The inset in (a) shows the schematic configuration of transport measurements. An oscillating demagnetization process was performed to eliminate the residual field in the measurement system. The Resistances were measured in cooling mode. Cooling measurements (a, b) were performed from the paramagnetic (PM) state.

It is necessary to study the corresponding magnetic field dependences of $R_{xx}$ and $R_{yx}$ at different temperatures. As shown by the green curve in Fig. 2(a1, a2), the initial state at -250 Oe is a single domain state with a negative magnetization. As the negative field decreases, the reverse domains begin to nucleate at the field $B_n$ (orange dotted line); thus, $R_{yx}$ begins to decrease (unsaturated) and the $\Delta R_{xx}$ component appears simultaneously. At the positive depinning field ($B_d$) (green dotted line), depinning of the domain walls occurs suddenly and $R_{yx}$ switches subsequently, leaving a small quantity of residual domains with negative magnetization. Simultaneously, $\Delta R_{xx}$ drops to a negative value that is close to zero. Further increasing the field to saturation $B_s$ (purple dotted line), the sample completely enters single domain state, with $R_{yx}$ becoming saturated and $\Delta R_{xx}$ disappearing. During the next field sweeping from 250 to -250 Oe, an antisymmetric behavior was observed. At different temperatures, similar antisymmetric loops are observed during the domain evolution (Fig. 2(a3, a4)). Figure 2 shows that the $\Delta R_{xx}$ component is antisymmetric in both magnetic field and magnetization, that is, $\Delta R_{xx}(M, B)=-\Delta R_{xx}(-M, -B)$, which is consistent with $R_{yx}$. Only $B$ and $M$ together can uniquely determine the domain states and related $\Delta R_{xx}$ for hard magnets.

The temperature-magnetic field diagram of $\Delta R_{xx}$ measured in field scanning from -250 to 250 Oe of $Co_3Sn_2S_2$ was summarized in Fig. 2(b), with the critical magnetic fields of $B_n$, $B_d$ and $B_s$ are indicated in Fig. 2(a1-a4). The nucleation field $B_n$ and depinning field $B_d$ increase with decreasing temperature. In the single-domain regime with either negative or positive saturation magnetization, $\Delta R_{xx}$ is absent. The $\Delta R_{xx}$ component emerges only in multi-domain states, which can therefore be co-modulated by the temperature and magnetic field.

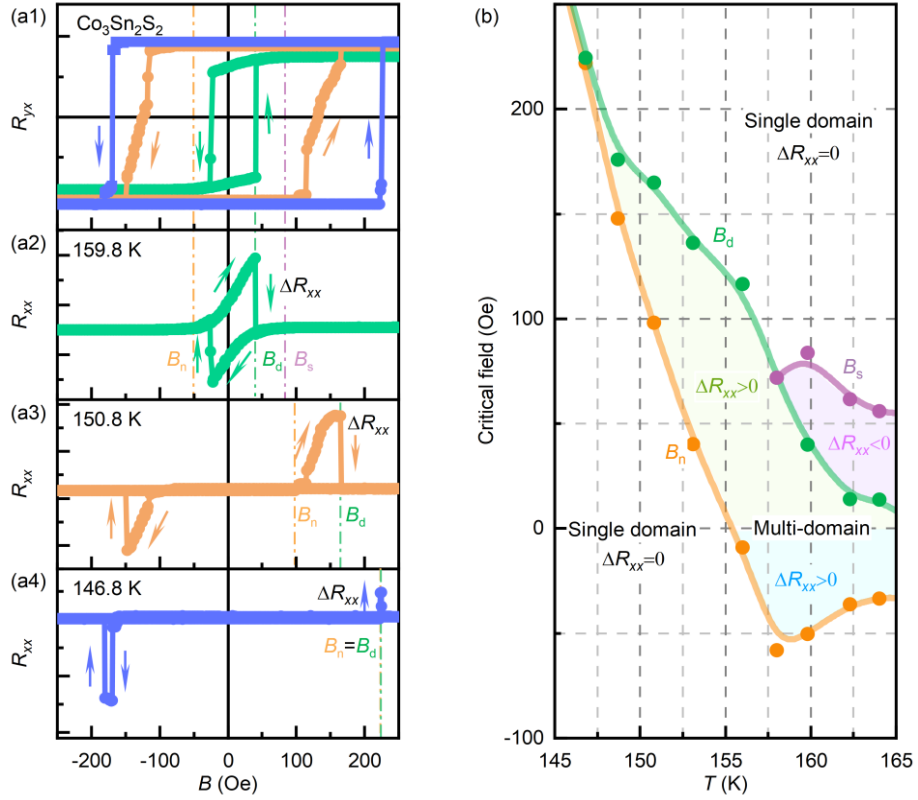

FIG. 2. (a1-a4) Magnetic field dependences of $R_{xx}$ and $R_{yx}$ of $Co_3Sn_2S_2$ nanoflake at different temperatures. (b) Temperature-magnetic field diagram of $\Delta R_{xx}$ measured in field scanning from -250 to 250 Oe. The multi-domain regime is surrounded by curves of critical magnetic fields of $B_n$, $B_d$ and $B_s$, corresponding to the dashed lines in (a). $B_n$: nucleation field; $B_d$: depinning field; and $B_s$: saturation field.

Although the antisymmetric magnetoresistance after magnetization saturating [9] and bowtie-like resistance below the coercivity field [15] have been observed in $Co_3Sn_2S_2$, our observations during domain evolution are quite different. To elaborate our results and obtain a more general understanding, we propose a concise and domain model, as depicted in Fig. 3. In a homogeneous sample with the single-domain state, the voltages between A-B and C-D electrodes are both determined by the longitudinal electric field ($E_x$) and distance ($L$), that is, $U_{AB}=U_A-U_B=U_{CD}=E_xL=U_0$. When an ideal striped magnetic domain wall or heterojunction interface is introduced in the middle of the device (blue line in Fig. 3(a1)), the AHE generates different electric fields at the two sides of the wall or interface. Considering the 180° domain wall as an example, hole carriers accumulate at C and B electrodes, while electron carriers accumulate at A and D. The anomalous Hall voltages between the transverse C-A and D-B electrodes exhibit opposite signs. Consequently, an additional voltage contribution originating from the anomalous Hall voltage can be introduced through the wall in the longitudinal direction along both device boundaries. The voltages of the longitudinal A-B and C-D electrodes are opposite in sign. More generally, the formulas are also applicable in the case of Fig. 3(a2), in which random multi-domain states including 180° or arbitrary-angle reverse magnetic domains, and inhomogeneous compositional domains are distributed in the matrix (see more details in Fig. S3).

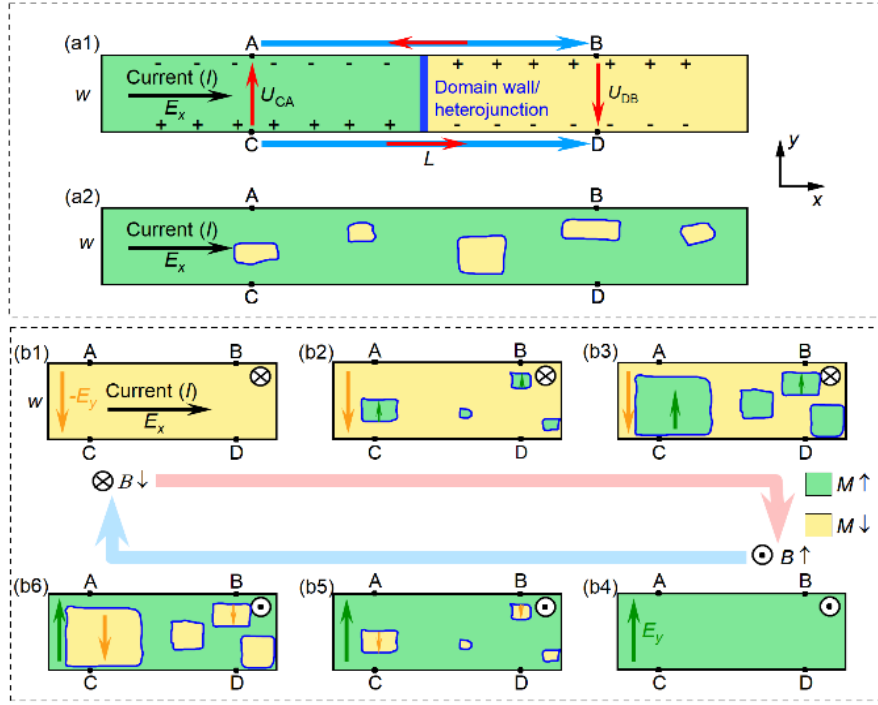

FIG. 3. (a) Schematic diagram of AHE induced anomalous longitudinal voltage based on ideal stripe magnetic domains or heterojunction. (a2) Schematic of the multi-domain state. (b) The evolution of the domain with the magnetic field. The blue and red arrows correspond to the directions of sweeping field. In all measurements, the current and magnetic field are along the $x$-axis and $z$-axis, respectively. The length between A(C) and B(D) is $L$, and the sample width is $w$.

As a typical process, the voltage evolution with the magnetic fields in case of multiple 180° magnetic domains is depicted in Fig. 3(b). In the single domain state with a negative magnetization (Fig. 3(b1)), $U_{AB}=U_{CD}=U_0$, with an anomalous Hall electric field of $-E_y$. As the negative magnetic field decreases, reverse domains with various sizes randomly appear in the matrix (Fig. 3(b2)), and some of them are located between C and A or D and B electrodes. The sign of the electric field in the reverse domains is opposite to that of the matrix. $U_{CA}$ ($U_{DB}$) thus decreases and $\Delta U_{AB}$ ($\Delta U_{CD}$) emerges, which is consistent with the measured decrease of saturated $R_{yx}$ and the appearance of $\Delta R_{xx}$ under the magnetic field $B_n$ in the $Co_3Sn_2S_2$ device (Fig. 2(a)). Then the reverse domains grow up with increasing the magnetic field (Fig. 3(b3, b4)), and $U_{CA}$ ($U_{DB}$) and $\Delta U_{CD}$ ($\Delta U_{AB}$) change subsequently. As the reverse domains tend to nucleate at the same locations of defects[16-19], $U_{CA}^{(b5)}=-U_{CA}^{(b2)}$ and $U_{DB}^{(b5)}=-U_{DB}^{(b2)}$, and $\Delta U_{CD}^{(b5)}=-\Delta U_{AB}^{(b5)}=-\Delta U_{CD}^{(b2)}=\Delta U_{AB}^{(b2)}$. This results in the antisymmetric longitudinal voltage $\Delta U_{CD}(M, B)=-\Delta U_{CD}(-M, -B)$, which is consistent with $\Delta R_{xx}(M, B)=-\Delta R_{xx}(-M, -B)$ observed in Fig. 2(a). From this multi-domain picture, the $\Delta U$ effect can be observed when there is a difference in the voltages induced by reverse domains with different widths located between two pairs (CA and DB) of Hall electrodes, which is significantly different from the single domain-wall picture (Fig. 3(a1)). For the heterojunction case, $\Delta U_{CD}$ ($\Delta U_{AB}$) is also antisymmetric in both $M$ and $B$ as this effect is induced by the AHE. However, in these devices, the sizes of magnetic domains can be modulated by external fields, whereas the sizes of heterojunction and compositional domains are stable.

Based on the analysis in Fig. 3, the longitudinal voltage is not simply equal to $U_0$ because of the unequal anomalous Hall voltages on both sides of the domain wall/heterojunction. This is an electric field re-distribution induced by the AHE, rather than a real change in the longitudinal resistance of the materials. To eliminate the influence of current ($I$), we use $R^*=\Delta U/I$ to represent this effect, termed domain-wall Hall magnetoresistance ($R^*$) in this work. In general, this Hall magnetoresistance can be induced by any nonsymmetric transverse voltage between two pairs of Hall electrodes within any multi-domain, such as different magnetic-domain distributions, inhomogeneous compositional domains, or heterogeneous junctions.

Four fundamental rules of $R^*$ can be revealed as follows: First, $R^*$ only appears in the multi-domain states ($U_{CA} \neq U_{DB}$) and should be considered particularly in the magnetization process of materials with a large AHE. Second, $R^*_{CD}=-R^*_{AB}$, which is a way to distinguish between $R^*$ and normal $R$ by examining the signs measured from two boundaries of the sample. Third, $R^*(M, B)=-R^*(-M, -B)$, with the same field symmetry as the anomalous Hall resistance. More importantly, the fourth rule states that $R^*$ is related to the transverse AHE between A (B) and C (D) electrodes. Therefore, $R^*$ originates from the Berry-phase-driven AHE in magnetic Weyl semimetals with a topology enhancement effect, which is quite different from the conventional magnetic materials. The concise formula provides a clear relationship between the longitudinal pseudo-resistance and transverse Hall resistance, even without knowing the sizes and numbers of domains in devices. The second and third rules have been addressed in conventional Co/Pt multilayers with a single domain wall [20,21], magnetic two-dimensional materials with a step edge to construct the domain wall [22,23], Co-Tb films with tilted domain walls [24], and wedge samples without domain walls [25]. In these studies, this phenomenon was referred to as antisymmetric magnetoresistance. These studies provided fundamental understanding on the similar magnetic field-dependent behaviors in ferromagnetic state. However, in our study, $R^*$ occurs in multi-domain wall states. Additionally, the numerical solution based on continuously changing domain walls with respect to magnetic field and temperature is highly challenging. Through our concise calculations, we provide an effective formula that directly relates $R^*$ to anomalous Hall voltage in multi-domain walls cases, which simplifies the practical problems.

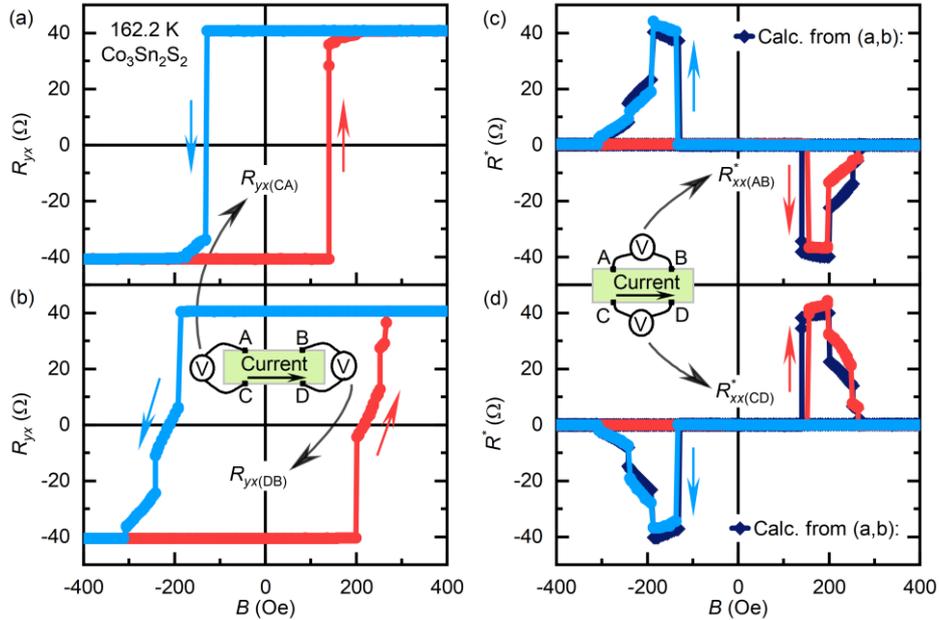

FIG. 4. (a)-(d) Magnetic field-dependent $R_{yx(CA)}$, $R_{yx(DB)}$, $R^*_{xx(AB)}$ and $R^*_{xx(CD)}$ of the $Co_3Sn_2S_2$ nanoflake with the thickness of 17.4 nm. To eliminate the effect caused by the mismatch of electrodes, the formula $R_{yx}=R_{yx}$(measure)-$(R_{yx}(+M, B=0)-R_{yx}(-M, B=0))/2$ was employed to obtain (a) and (b). The equation $R^*=R_{xx}(B)-R_{xx}(0)$ was used to obtain (c) and (d). (a) and (b) were measured simultaneously, (c) and (d) were measured simultaneously.

We further validate the above rules by measuring $R_{yx(CA)}$, $R_{yx(DB)}$, $R_{xx(CD)}$ and $R_{xx(AB)}$ of pristine $Co_3Sn_2S_2$ with the thickness of 17.4 nm. The antisymmetric relation of anomalous Hall resistance $R_{yx}(M, B)=-R_{yx}(-M,-B)$ in Fig. 4(a, b) implies that the reverse domains are always nucleated at the same defects during magnetization in the device. For the $R_{yx(CA)}$ curves in Fig. 4(a), only one step of sharp switching of the domain is observed, whereas for $R_{yx(DB)}$ in Fig. 4(b), three steps are observed, indicating that the size and number of reverse domains between CA electrodes are different from those of DB. This subsequently results in the $R^*$ effect, according to the picture in Fig. 3. By using the formula to obtain $R^*_{xx(AB)}$ and $R^*_{xx(CD)}$. They closely coincide with our experimental measurements (Fig. 4(c, d)), which proves good verification of our conclusions. And multi-resistance state was realized in this sample. The antisymmetric relationship $R^*(M, B)=-R^*(-M, -B)$ deduced in Fig. 3 is well confirmed by our experiments in Fig. 4(c, d).

Our study shows that the Hall magnetoresistance of different magnitudes can be obtained and tuned. Compared with previous reports on conventional magnetic materials [20,22-24], $R^*$ improves by an order of magnitude in the well-defined magnetic Weyl semimetal $Co_3Sn_2S_2$ (Table 1), which is important for multi-resistance-state modulation for data storage and logic circuits. As a significant parameter, $R^*$ is directly related to the topology effect of Weyl electrons, which indicates the importance of topology physics for future spintronic materials.

Table 1 Comparison of $R^*$ induced by AHE with reported materials and our $Co_3Sn_2S_2$ nanoflakes.

| Material | $R^*(\Omega)$ |
|---|---|
| Co/Pt [20] | 0.09 |
| $Fe_3GeTe_2$ [22] | 1.20 |
| $Co_xTb_{1-x}$ [24] | 0.77 |
| $Fe_3GaTe_2$ [23] | 1.95 |
| $Co_3Sn_2S_2$-this work | 50.80 |

In summary, we study the intriguing longitudinal transport behavior of magnetic Weyl semimetal $Co_3Sn_2S_2$ nanoflakes. Giant domain-wall Hall magnetoresistance related to the multi-domains was observed, exhibiting clear antisymmetry with respect to $M$ and $B$. We proposed a multi-domain model to understand this significant transport, which indicated that the AHE-carrying multi-domains that lead to the Hall magnetoresistance, not the intrinsic change in resistance. The fundamental rules of this pseudo-resistance were revealed and experimentally verified. The topologically-enhanced Berry phase in $Co_3Sn_2S_2$ elevated the Hall magnetoresistance by an order of magnitude. This work provides a universal understanding of longitudinal transport in the multi-domain states of magnetic Weyl semimetals, providing a giant Hall magnetoresistance effect for multi-resistance-state modulation in future spintronics applications.